\documentclass[prl,twocolumn,showpacs,preprintnumbers,amsmath,amssymb,superscriptaddress]{revtex4}

\usepackage{graphicx}
\usepackage{dcolumn}
\usepackage{bm}
\usepackage{amsmath}
\usepackage{amsfonts}
\usepackage{amssymb}
\usepackage{color}
\usepackage{epsfig,float,afterpage,amssymb,wrapfig,psfrag}

\begin{document}
\title{Universal out-of-equilibrium transport in Kondo-correlated quantum dots: renormalized dual fermions on the Keldysh contour}
\author{Enrique Mu\~{n}oz}
\affiliation{Facultad de F\'isica, Pontificia Universidad Cat\'olica de Chile, Casilla 306, Santiago 22, Chile.}
\author{C.~J.~Bolech}
\affiliation{Department of Physics, University of Cincinnati, Cincinnati, Ohio 45221-0011, United States.}
\author{Stefan Kirchner}
\affiliation{Max Planck Institute for the Physics of Complex Systems,
01187 Dresden, Germany.}
\affiliation{Max Planck Institute for Chemical Physics of Solids,
01187 Dresden, Germany.}

\date{\today}

\begin{abstract}
The non-linear conductance of semiconductor heterostructures and single molecule devices exhibiting
Kondo physics has recently attracted attention. We address the observed sample-dependence of the measured steady state transport coefficients
by considering additional electronic contributions in the effective low-energy model underlying these experiments
that are absent in particle-hole symmetric setups.
A novel version of the superperturbation theory of Hafermann et al. in terms of dual fermions is developed,
which
correctly captures the low-temperature behavior.
We compare our results with the measured transport coefficients.
%
\end{abstract}



\maketitle

Quantum matter out of equilibrium is currently investigated in a wide range of settings ranging from cold atom setups and light-matter systems
to various condensed matter systems.  Depending on the context, the focus ranges
from thermalization of quantum matter to the description of relaxation processes to the microscopic characterization of non-thermal
steady states.  In condensed matter systems, with couplings to well-defined  heat- and particle reservoirs, current-carrying steady states are of particular interest \cite{Bonca,Hewson}.\\
In this letter, we are concerned with the nonlinear conductance of a model system of strong electron-electron interaction.
Traditionally, the calculation of transport properties
 on the basis of the fluctuation-dissipation theorem is fairly well developed.
Yet, no generally valid method  exists to go beyond the linear-response regime, as {\it e.g.} a Boltzmann equation based approach relies on
well-defined quasi-particles and relaxation-time-like approximations.
Of particular current interest is therefore the effect of strong electron-electron correlation on electrical and thermal conductivities beyond the linear response regime.
Kondo-correlated quantum dots
have served as ideal model systems to address  this interplay between out-of-equilibrium dynamics and strong correlations both experimentally and theoretically.
In equilibrium, the Kondo effect leads to an enhancement
of the linear conductance $G=dI/dV|_{V=0}$  to close to twice the quantum of conductance at sufficiently low temperatures ($I$ is the current through the quantum dot and $V$ the applied bias voltage) independent of any details of {\it e.g.} the density of states of the leads.
The fate of this universality away from equilibrium has been  subject of intense research~\cite{Schiller.95,Kaminski.00,Oguri.01,Konik.02,Hewson.05}.
Recently, the universal aspects of steady-state charge transport in the Kondo regime beyond  linear response through semiconductor heterostructures and various single molecule devices  have been addressed experimentally~\cite{Grobis.08,Scott.09,Kretinin.11}. It was found that the prefactors $\alpha$ and $\gamma$ of the non-linear conductance, defined via ($k_B=1$)
\begin{equation}
(G_0\!\!-\!\!G(T,V))/(c_TG_0)\!\!=\!\!\Big(\dfrac{T}{T_K}\Big)^2\!\!+\alpha \Big(\dfrac{eV}{T_K}\Big)^2\!\!-\gamma c_T \Big(\dfrac{eV\,T}{T^2_K}\Big)^2
\end{equation}
differ significantly across different classes of devices.
Here, $T$ is temperature, $G_0=G(T\rightarrow 0,V=0)$ and $T_K$ is a dynamically generated low energy scale, {\it i.e.} the Kondo temperature.

Our primary motivation is to address the systematic difference between the results reported in~\cite{Grobis.08} ($\alpha_{G}=0.1$, $\gamma_G=0.5$) and~\cite{Scott.09} ($\alpha_S=0.05$, $\gamma_S=0.1$) within the single-level Anderson impurity model (SIAM) as the effective low-energy model for these devices.
In the strong coupling regime, this model is equivalent to the Kondo model plus a potential scattering term generated away from particle-hole (p-h) symmetry. Particle-hole symmetry can 
easily be  broken either locally on the device itself (see below Eq.~(\ref{eq:Hamiltonian})) 
or in the leads connected to the device (see below Eq.~(\ref{eq:DOS})~\cite{note2}).
Consequently,
realistic devices are generically not p-h symmetric and it is important to understand the effect of p-h asymmetry on transport properties. An immediate consequence of p-h asymmetry is that the number of electrons localized on the device is no longer fixed to be 1/2 (per spin component).\\
Theoretically, not much is known about $\alpha$ and $\gamma$.
A full solution of the
SIAM out of equilibrium is not available and the  calculation of these transport coefficients is challenging.
Results for $\alpha$ obtained from exactly solvable cases are not directly applicable~\cite{Schiller.95,Majumdar.98}. Standard approaches, {\it e.g.} the numerical renormalization group (NRG) yield only linear response transport coefficients~\cite{Costi.94c}.
Selfconsistent methods can in principle be extended to the non-linear response regime. They are conserving by construction~\cite{Baym.61} but either fail to capture the correct ground state as {\it e.g.}~the non-crossing approximation or the extension onto the Keldysh contour is too involved~\cite{Kirchner.02}.
As the potential scatterer is a marginally irrelevant perturbation  it is expected to modify the transport coefficients but its effect should be pertubatively accessible.
At p-h symmetry $\alpha\approx0.15$ has been obtained independently of
the amount of asymmetry in the lead-dot coupling between the two leads~\cite{Oguri.01,Oguri.05,Rincon.09,Sela.09,note1}.

The SIAM Hamiltonian is
$\hat{H} = \hat{H}_c + \hat{H}_d + \hat{H}_{d-c}$,
where
\begin{eqnarray}
\label{eq:Hamiltonian}
\hat{H}_c \!\!\!&&= \sum_{\lambda = L, R}
\sum_{k,\sigma}\epsilon_{k\lambda}
\hat{c}_{k\lambda\sigma}^{\dagger}\hat{c}_{k\lambda\sigma} \\
\hat{H}_d \!\!\!&&= \sum_{\sigma} E_d \hat{d}_{\sigma}^{\dagger}\hat{d}_{\sigma}
+U\left(\hat{d}_{\uparrow}^{\dagger}\hat{d}_{\uparrow} - \frac{1}{2}\right)\left(\hat{d}_{\downarrow}^{\dagger}\hat{d}_{\downarrow} - \frac{1}{2}\right) - \frac{U}{4}
\nonumber\\
\hat{H}_{d-c} \!\!\!&&= \sum_{\lambda = L,
R}\sum_{k,\sigma}\left(V_{k\lambda}\hat{d}_{\sigma}^{\dagger}\hat{c}_{k\lambda\sigma}
+V_{k\lambda}^{*}\hat{c}_{k\lambda\sigma}^{\dagger}\hat{d}_{\sigma}\right).\nonumber
\end{eqnarray}
Here, $\hat{H}_c$ is the Hamiltonian for electrons in the metallic leads $\lambda=L$ and $\lambda=R$. 
$\hat{H}_{d}$ describes the localized states in the dot, including the Coulomb interaction, and $\hat{H}_{d-c}$ is the coupling term between the
dot and the leads. We have defined $E_d = \epsilon_d + U/2$. For the p-h symmetric case $\epsilon_{d} = -U/2$
and hence $E_{d} = 0$.

Beyond setting up a systematic expansion for $\alpha$ and $\gamma$ in terms of $E_{d}$ and up to O($V^2$),
 we also address the issue of current conservation beyond O($V^2$).
Away from p-h symmetry, a proper treatment of the (renormalized) interaction vertex is necessary to reproduce
 {\it e.g.} the correct local occupation already in equilibrium. Since, by continuity, particle flow is connected 
to the rate of change of the local occupation, any sensible approximation has to respect 
the corresponding symmetries of the interaction vertex in order to be current-conserving~\cite{Baym.61}. 
As 
discussed by Hershfield et al.~\cite{Hershfield.92}, for perturbation theory in $U$, steady state current 
conservation holds only in the p-h symmetric SIAM.
We therefore develop an approach to transport in the p-h asymmetric SIAM based on
 dual fermions~\cite{Rubtsov.08} 
that is based 
on perturbation theory in $U$ for the p-h symmetric SIAM~\cite{Hewson.93,Oguri.01,Zlatic.83,Yamada.79}.
As demonstrated explicitly, our results are rigorous up to O($V^2$)
and are current conserving (beyond O($V^2$)).
As the p-h symmetric SIAM is interacting, the expansion around it is delicate. 
We use the dual fermion method~\cite{Hafermann.09,Jung.11} which yields a formal expansion built around the 
4-point vertex of the
reference system with $E_{d} = 0$.
This systematically extends the work of Yamada and Yosida and Zlati\'{c} and Horvati\'{c} to the 
asymmetric SIAM~\cite{Yosida.70,Yamada.75,Yamada.79,Zlatic.83,Horvatic.87} and results in
a controlled expansion for the transport coefficients up to, including, $O(U^2 E_d^2)$. 
A generalization to higher orders is possible~\cite{note3}.


The generating functional on the Keldysh contour is given by
\begin{eqnarray}
Z = \int\mathcal{D}[\hat{\psi}^{\dagger},\hat{\psi}]\mathcal{D}[\hat{\Phi}^{\dagger},\hat{\Phi}]e^{iS[\hat{\psi}^{\dagger},\hat{\psi},\hat{\Phi}^{\dagger},\hat{\Phi}]},
\label{Eq:partitionfunction}
\end{eqnarray}
where the action on the Keldysh contour is expressed in terms
of a functional integral over time-dependent Grassmann fields, $\hat{\psi}_{k\lambda\sigma}^{\dagger}(t) = \left(c_{k\lambda\sigma}^{-}(t), c_{k\lambda\sigma}^{+}(t) \right)^\dagger$ and $\hat{\Phi}^{\dagger}(t) = \left(d_{\sigma}^{-}(t), d_{\sigma}^{+}(t) \right)^\dagger$. Here, the indices $\pm$ refer to the
time-ordered (-) and anti-time-ordered (+) path along the closed Keldysh contour.
Each lead ($L/R$) is taken to be in equilibrium and characterized by its temperature ($T_L=T_R=T$) and its chemical potential ($\mu_L/\mu_R$).\\
The lead electrons are non-interacting and the resulting Gaussian integrals can be carried out, resulting in
\begin{eqnarray}
Z = \int\mathcal{D}[\hat{\Phi}^{\dagger}_{\sigma\omega},\hat{\Phi}_{\sigma\omega}]e^{iS[\hat{\Phi}^{\dagger}_{\sigma\omega},\hat{\Phi}_{\sigma\omega}]}~,
\label{eq05}
\end{eqnarray}
where the effective action $S$ is given by
\begin{eqnarray}
 S[\hat{\Phi}^{\dagger}_{\sigma\omega},\hat{\Phi}_{\sigma\omega}] &=&  S_{U}[\hat{\Phi}^{\dagger}_{\sigma\omega},\hat{\Phi}_{\sigma\omega}]\nonumber \\
&-& \int_{-\infty}^{+\infty}\frac{d\omega}{2\pi}\sum_{\sigma}\hat{\Phi}^{\dagger}_{\sigma\omega}E_{d}\hat{\sigma}_{3}\hat{\Phi}_{\sigma\omega}~,
\label{eq:effA}
\end{eqnarray}
and
\begin{eqnarray}
\label{eq07}
&&  S_{U}[\hat{\Phi}^{\dagger}_{\sigma\omega},\hat{\Phi}_{\sigma\omega}] =  S_{U}^{int}[\hat{\Phi}^{\dagger}_{\sigma\omega},\hat{\Phi}_{\sigma\omega}] \\
&& + \int_{-\infty}^{+\infty}\frac{d\omega}{2\pi}
\sum_{\sigma}\hat{\Phi}^{\dagger}_{\sigma\omega}(\omega + (\Gamma_{L} + \Gamma_{R}))\hat{\sigma}_{3}\hat{\Phi}_{\sigma\omega}~ \nonumber
\end{eqnarray}
is the effective action for a p-h symmetric ($E_{d} =0$) and interacting ($U \ne 0$) system.
Here,
\begin{eqnarray}
\label{eq:DOS}
\Gamma_{\lambda} = -\sum_{k,\sigma}\frac{|V_{k\lambda}|^{2}}{\omega - \epsilon_{k\lambda} + i\eta^{+}}\,\,\,\,\,\,\,\rm{for}\,\,\, \lambda = L, R.
\end{eqnarray}
For simplicity, we assume that the density of states of left and right lead, $\rho_\lambda(\omega)=\sum_{k}\delta(\omega-\epsilon_{k\lambda})$, are identical and p-h
symmetric $\rho_{\lambda}(\omega)=\rho_{\lambda}(-\omega)$~\cite{note2}.
In the wide band limit, we set $i\Delta=\Gamma_L+\Gamma_R$.
To generate an expansion in terms of $E_d$, we decouple the 2nd term on the RHS of Eq.(\ref{eq:effA}) into
$\phi^\dagger_{\sigma\omega} {{\mathbf g}}^{-1}_{\sigma\omega}\hat{\Phi}_{\sigma\omega}$ via a
fermionic Hubbard-Stratonovich transformation, where
$\mathbf{g}_{\sigma,\omega}$ is the
Green's function for the interacting ($U\ne 0$) and symmetric ($E_{d}=0$) SIAM~\cite{Rubtsov.08}.
One can show~\cite{note3,Rubtsov.08}
\begin{eqnarray}
\mathbf{G}_{\sigma,\omega} = -E_{d}^{-1}\hat{\sigma}_{3} +
\left(\mathbf{g}_{\sigma,\omega}E_{d}\hat{\sigma}_{3}\right)^{-1}
\mathbf{G}_{\sigma,\omega}^{f}\left(E_{d}\hat{\sigma}_{3}\mathbf{g}_{\sigma,\omega}\right)^{-1},
\label{eq1}
\end{eqnarray}
where $\mathbf{G}_{\sigma,\omega}$ is the Green's function
matrix for the interacting ($U\ne 0$) asymmetric ($E_{d}\ne 0$)
SIAM, $\hat{\sigma}_{3}$ is the third Pauli matrix, and
$\mathbf{G}_{\sigma,\omega}^{f}$ is the dual
fermion matrix Green's function, obtained from the solution of the matrix Dyson
equation
\begin{eqnarray}
\mathbf{G}_{\sigma,\omega}^{f} = \mathbf{G}_{\sigma,\omega}^{f(0)}
+ \mathbf{G}_{\sigma,\omega}^{f(0)}\mathbf{\Sigma}_{\sigma,\omega}^{f}
\mathbf{G}_{\sigma,\omega}^{f},
\end{eqnarray}
where the bare dual fermion Green's function is defined by
$\mathbf{G}_{\sigma,\omega}^{f(0)} =
-\mathbf{g}_{\sigma,\omega}\left(\mathbf{g}_{\sigma,\omega} -
E_{d}^{-1}\hat{\sigma}_{3}\right)^{-1}
\mathbf{g}_{\sigma,\omega}$.
The dual fermion selfenergy $\mathbf{\Sigma}_{\sigma,\omega}^{f}$ is given in terms
of $\mathbf{g}_{\sigma,\omega}$ and the 4-point vertex of the interacting ($U\ne 0$) and symmetric
($E_{d}=0$) SIAM~\cite{note3}. So far, no approximation has been made and this expansion is expected 
to work for small as well as large $E_d$~\cite{Hafermann.09}.
We proceed by solving the reference system ($E_d=0$) within the renormalized perturbation theory around the strong 
coupling fixed point~\cite{Yamada.75,Hewson.93,Oguri.01}.
For a systematic expansion in $E_{d}$ up to O($E_d^2$), we keep 
only the first two terms in the Dyson series for $\mathbf{G}_{\sigma,\omega}^{f(0)}$.
As a result,
the explicit expression for the retarded self-energy at finite bias voltage $\mu_L-\mu_R=eV$ obtained from
our superperturbation scheme up to, including, $O(T^{2}V^{2})$, is~\cite{note3,note4}
\begin{eqnarray}
\label{eqA1}
&&\Sigma_{E_{d}}^{r}=(1-\tilde{\chi}_{++})\omega + E_{d} -
\tilde{\chi}_{++}^{-1}E_{d}\left(\frac{U}{\pi\Delta}\right)\left\{
1 - \frac{\tilde{\chi}_{++}^{2}}{3}\right.\nonumber\\
&&\times\left.\left[\left(\frac{\pi T}{\Delta} \right)^{2} + \zeta\left(\frac{eV}{\Delta}\right)^{2} \right]
+7\frac{\zeta}{9}\tilde{\chi}_{++}^{4}\left(\frac{\pi T eV}{\Delta^{2}} \right)^{2}\right\}\nonumber \\
&&-i\frac{\Delta}{2}\left(\frac{U}{\pi\Delta} \right)^{2}\\
&&\times \left[
\left(\frac{\omega}{\Delta} \right)^{2} + \left(\frac{\pi T}{\Delta}\right)^{2} + \zeta\left(\frac{e V}{\Delta} \right)^{2}-\frac{\zeta}{3}\left(\frac{\pi T e V}{\Delta^2}\right)^2\tilde{\chi}_{++}^{2}\right],\nonumber
\end{eqnarray}
with $\tilde{\chi}_{++} = 1 + (3 - \pi^{2}/4)(U/\pi\Delta)^{2} + O(U^{4})$~\cite{Hewson.93,Yamada.79}
and $\zeta=3\kappa/(1+\kappa)^2$ where $\kappa=\Gamma_L/\Gamma_R$ measures the asymmetry in the lead-to-dot couplings.
Notice that there are no terms of $O(E_{d}^{2}U)$ nor $O(E_{d}^{2}U^{2})$ in Eq.(\ref{eqA1}). The next
leading correction to the retarded selfenergy is $O(E_{d}^{3}U,E_{d}U^{3})$~\cite{note3}.
For $U=0$, Eq.~(\ref{eqA1}) reduces to the corresponding result of the resonant level model.

We now turn to a discussion of the current.
The steady-state current through the dot~\cite{Hershfield.92,Meir.92},
\begin{equation}
I \!=\! \left(\frac{ e}{\hbar}\right)\int_{-\infty}^{+\infty}\!\!\!d\omega
\frac{4\Gamma_{R}\Gamma_{L}}{\Gamma_{R}+\Gamma_{L}}[f_{L}(\omega) - f_{R}(\omega)] A(\omega,T,V),
\label{eq:current}
\end{equation}
follows from the continuity equation and relies on current conservation $I_L+I_R=0$ in the steady state to 
recast $I$ entirely in terms of the spectral density.
As a result, Eq.~(\ref{eq:current}) poses a strong constraint on admissible local distribution functions
$F(\omega,T,V)$, where $F$ is defined through $G^{-+}=F(\omega,T,V)(G^a-G^r)$~\cite{note3}.
Here, $I_{L/R}$ is the current from the left/right lead to the dot,
$A(\omega,T,V)$ is the local spectral density (in the presence of the dot-lead coupling) and $f_L$/$f_R$ is 
Fermi function in the left/right lead, respectively. A second local distribution function $\tilde{F}(\omega,T,V)$ 
can be introduced via $\Sigma^{-+}=\tilde{F}(\omega,T,V)(\Sigma^a-\Sigma^r)$.
For the SIAM considered here one can show that $F(\omega,T,V)=\tilde{F}(\omega,T,V)$ in the steady state limit. 
This in turn implies $G^{-+}\Sigma^{+-}=G^{+-}\Sigma^{-+}$ which ensures current 
conservation~\cite{Hershfield.92,note3}.
Note, that in general one cannot conclude $F=\tilde{F}$ away from equilibrium. 

Current conservation of our approach beyond $O(V^2)$ 
follows from the general relations
 $\Sigma^{++}_{E_d}+\Sigma^{--}_{E_d}-\Sigma^{+-}_{E_d}-\Sigma^{-+}_{E_d}=0$, 
$\Sigma^{++}_{E_d}=-(\Sigma^{--}_{E_d})^*$, $\Sigma^r_{E_d}=\Sigma^{--}_{E_d}-\Sigma^{-+}_{E_d}$ and 
Eq.~(\ref{eqA1}) which imply
\begin{eqnarray}
\!\!\!\!&&\!\!\!\!F(\omega,T,V)(\Sigma_{E_{d}}^{a}- \Sigma_{E_{d}}^{r})=
i\Delta\left(\frac{U}{\pi\Delta}\right)^{2}\nonumber \\
\!\!\!\!&&\!\!\!\!\times\left[\left(\frac{\omega}{\Delta}
\right)^{2} + \left(\frac{\pi T}{\Delta} \right)^{2}  +  \zeta\left(\frac{eV}{\Delta}\right)^{2} -\frac{\zeta}{3}\left(\frac{\pi T e V}{\Delta^2}\right)^2 \tilde{\chi}^2_{++}\right]\nonumber \\
\!\!\!\!&&\!\!\!\!\times  f_{eff}(\omega,T,V)
=\Sigma_{E_{d}}^{-+},
\label{eq47}
\end{eqnarray}
where we introduced $f_{eff}(\omega,T,V)=(\kappa f_L+f_R)/(1+\kappa)$. Eq.~(\ref{eq47}) shows that
within our scheme $F(\omega,T,V)=\tilde{F}(\omega,T,V)$.
The local distribution function $F$ turns out to be~\cite{note3}
\begin{equation}
F(\omega,T,V)=\frac{\Gamma_Lf_L+\Gamma_R f_R -f_{eff}(\omega,T,V) \mbox{Im}\Sigma^r}{1-\mbox{Im}\Sigma^r}.
\end{equation}


The non-linear conductance follows from Eq.~(\ref{eq:current}) and the approximation for
$A(\omega,T,V) = -\pi^{-1}{\rm{Im}}\, G^{r}$,
where $G^{r}=(\omega+i\Delta-\Sigma_{E_d}^{r})^{-1}$ is the retarded Green function.
We are primarily interested in the transport coefficients in
the vicinity of the strong coupling fixed point, where our expansion is in terms of renormalized parameters~\cite{Hewson.93}.
The renormalized parameters are defined as $\tilde{\epsilon}_{d}=
E_{d}/\Delta$, $\tilde{\Delta} = \tilde{\chi}_{++}^{-1}\Delta$,
$\tilde{u}=\tilde{\chi}_{++}^{-1}(U/\pi\Delta)$.
In terms of these, one finds
\begin{multline}
\dfrac{G(T,0)-G(T,V)}{G_0}=c_V \left( \dfrac{eV}{\tilde{\Delta}}\right )^2 -c_{TV} \left( \dfrac{eV}{\tilde{\Delta}}\right )^2 \left( \dfrac{k_B T}{\tilde{\Delta}}\right )^2 \\
-c_{\mbox{\tiny $VE_d$}}\left( \dfrac{eV}{\tilde{\Delta}}\right ) + c_{\mbox{\tiny $TVE_d$}} \left( \dfrac{eV}{\tilde{\Delta}}\right )
\left( \dfrac{k_B T}{\tilde{\Delta}}\right )^2,
\label{eq:cond}
\end{multline}
where
\begin{eqnarray}
\label{eq:cT}
G(T,V=0)&=&G_0\left[1-c_T\left( \dfrac{k_B T}{\tilde{\Delta}}\right )^2\right],\\
c_T&=&\frac{\pi^{2}}{3}\frac{1+2\tilde{u}^{2}+\tilde{\epsilon}_{d}^{2}
[(8-5\tilde{u})\tilde{u}-3]}{\left(1 + (1-\tilde{u})^{2}\tilde{\epsilon}_{d}^{2} \right)^{2}}
\nonumber.
\label{eq:cT2}
\end{eqnarray}
The zero-temperature linear conductance
$G_0=(2e^2/h)\frac{4\zeta}{3}(1+(1-\tilde{u})^{2}\tilde{\epsilon}^2_d)^{-1}$ reproduces the exact result from
Friedel's sum rule up to $O(\tilde{u}^{2}\tilde{\epsilon}^2_d)$ as
$\sin^{2}(\pi n_{d})
\sim 1 - (1-\tilde{u})^{2}\tilde{\epsilon}_{d}^{2}$, for $n_{d}$ the local occupation per spin component.
For the transport coefficients in Eq.~(\ref{eq:cond}), we find
\begin{eqnarray}
c_{V} &=& 1 + \frac{\tilde{u}^{2}}{2} -\zeta\left(1 - \tilde{u}^{2} \right)
-\tilde{\epsilon}_{d}^{2}\left(1 - \tilde{u} \right)\nonumber \\
&&\times H_{V}(\tilde{u},\zeta) + O(\tilde{\epsilon}_{d}^{4})\\
\label{eq:ctv}
c_{TV} &=& \pi ^2\left[2 (1 - \zeta)
+ \frac{\tilde{u}^2}{2} (9 - 5\zeta) \right]
-\tilde{\epsilon}_{d}^{2}\left(1 - \tilde{u} \right)\nonumber \\
&&\times H_{TV}(\tilde{u},\zeta) + O(\tilde{\epsilon}_{d}^{4}) \\
\label{eq:cved}
c_{VE_{d}} &=& 2\left(\frac{1-\kappa}{1+\kappa}\right)(1-\tilde{u})\tilde{\epsilon}_{d} + O(\tilde{\epsilon}_{d}^{3})\\
\label{eq:cTved}
c_{TV E_{d}} &=& -2 \pi^{2}\left(\frac{1-\kappa}{1 +\kappa}\right)\left(2 + 3\tilde{u}^{2} \right)\nonumber \\
&&\times \left(1 -\tilde{u}\right)\tilde{\epsilon}_{d} + O(\tilde{\epsilon}_{d}^{3})
\end{eqnarray}
where we have defined the functions
$H_{V}(\tilde{u},\zeta) = 5 - 5\tilde{u} + \tilde{u}^{2} - \zeta\left(5 - 3\tilde{u} - 2\tilde{u}^{2} \right)$ and
$H_{TV}(\tilde{u},\zeta) = \pi^{2}\left[28 - 16\tilde{u} + \frac{81}{2}\tilde{u}^{2}
-\zeta\left(28 - \frac{22}{3}\tilde{u} + \frac{76}{3}\tilde{u}^{2} \right) \right]$.

\begin{figure}[t!]
 \centering
 \includegraphics[width=0.4\textwidth]{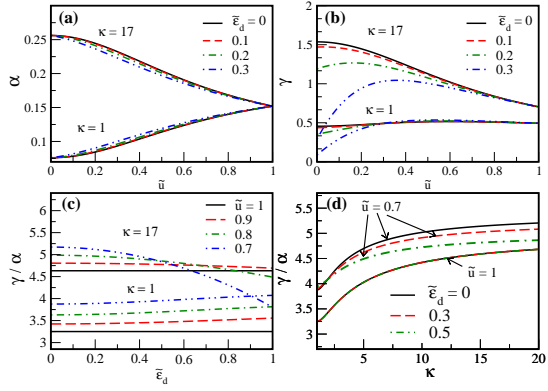}
 \caption{(Color online) Coefficients $\alpha$ (upper set) and $\gamma$ (lower set) versus the 
Different degrees of lead-to-dot asymmetry coupling: $\kappa = 20$ (left)
and $\kappa = 2$ (right) are compared, for different values of
particle-hole asymmetry $\tilde{\epsilon}_d$.}
 \label{fig:1}
\end{figure}

In Fig.~\ref{fig:1}, we show our results for $\alpha$ and $\gamma$ for various cuts through parameter space ($\tilde{u},\tilde{\epsilon}_d,\kappa$).
Note, that in the strong coupling limit ($\tilde{u}\rightarrow 1$) the dependence on $\tilde{\epsilon}_d$
vanishes reflecting the fact that this limit is p-h symmetric (see Fig.~\ref{fig:1}(a) and (b)).
$\gamma$ retains its dependence on $\kappa$  in this limit while $\alpha$ becomes independent of $\kappa$ for $\tilde{u}\rightarrow 1$. Fig.~\ref{fig:1}(c) and (d) show the ratio $\gamma/\alpha$.
According to Eqs.~(\ref{eq:cved}) and (\ref{eq:cTved}) $c_{\mbox{\tiny $VE_d$}}$ and $c_{\mbox{\tiny $TVE_d$}}$ are  proportional to the product of lead-dot asymmetry $\kappa$ and p-h asymmetry $\tilde{\epsilon}_d$ and hence may be small in most experimental realizations.
For the p-h symmetric case our expressions reduce to the results of Oguri and others~\cite{Oguri.01,Oguri.05,Rincon.09}.

\begin{figure}[t!]
 \centering
 \includegraphics[width=0.38\textwidth]{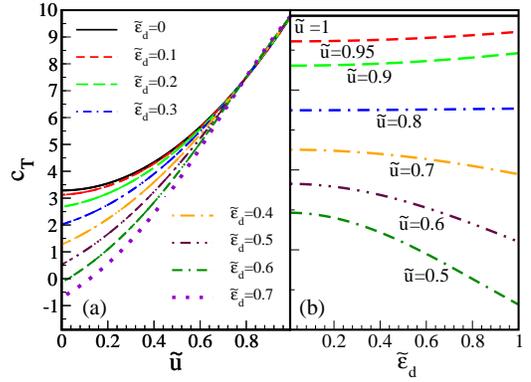}
 \caption{(Color online) Prefactor of $k_B^2T^2/\tilde{\Delta}^2$ of the linear conductance (a) vs. renormalized coupling strength (b) vs.p-h asymmetry.}
 \label{fig:2}
\end{figure}

We are now in a position to address the experimental results for $\alpha=c_V/c_T$ and 
$\gamma=c_{TV}/c_T^2$~\cite{Grobis.08,Scott.09}.
A major experimental challenge is to reliably extract the dynamically generated low-energy scale
$\tilde{\Delta}\sim T_K$ ($\tilde{\Delta}=4k_BT_K/\pi$ at $\tilde{u}=1$).
The phenomenological formula
$G(T,0)=G_0/(1+(2^{1/s}-1)(T/T_K)^2)^s$
is commonly employed to extract $T_K$~\cite{GoldhaberGordon.98}.
Evidently, the parameter $s$ fixes $c_T$  ($s=0.21$ as in~\cite{Grobis.08} leads to 
$c_T\approx 5.5$ and $s=0.22$~\cite{Scott.09} results in $c_T\approx 4.9$).
Eq.~(\ref{eq:cT2}) shows that $c_T$ is not only a function of $\tilde{u}$ but also depends on 
the p-h asymmetry through $\tilde{\epsilon}_d$, see Fig.~\ref{fig:2}. 
This complicates the experimental extraction of $T_K$.
In theory, $T_K$  is not unique away from p-h symmetry but will 
depend on the physical quantity used for its definition.

The reported values~\cite{Grobis.08,Scott.09} suggest that charge fluctuations are present in both experiments and the coefficients $c_{\mbox{\tiny $VE_d$}}$ and that $c_{\mbox{\tiny $TVE_d$}}$ are indeed vanishingly small.
Yet, they may have been detected in~\cite{Scott.09}.
The experimental values reported in \cite{Grobis.08} are compatible with {\it e.g.} $\tilde{u}=0.45,\tilde{\epsilon}_d=0.1,\kappa=1$ yielding $\alpha=0.1$ and $\gamma=0.51$.
While we can reproduce $\gamma_S$ of \cite{Scott.09}, it is not possible to
reproduce both consistently within the SIAM. The value $\alpha_S \sim 0.05$ is too small to be explained within the SIAM, as the minimum
value for $\alpha$ within the SIAM is $\alpha_{\mbox{\tiny min}} = 3/(4\pi^{2})\approx0.076$ (corresponding to $\tilde{\epsilon}_d=0$, $\tilde{u}=0$, $\kappa=1$). The underlying low-energy model of the experiment  \cite{Scott.09} can therefore not simply be the SIAM.  One possible generalization is that
more than one level participates in the low-energy properties. Then, already $G_0$ is no longer given solely
in terms of the occupation $n_d$ and the lead-to-dot couplings will enter explicitly~\cite{Kroha.03}.
A more likely alternative is that local phonon modes renormalize the transport coefficients $\alpha$ and $\gamma$ differently.

In summary, we have developed a novel analytic scheme based on dual fermions to obtain non-linear transport 
coefficients for the Anderson model. This approach gives a controlled expansion around the weak and strong 
coupling fixed points even away from particle-hole symmetry
and allows for a consistent calculation of charge and energy currents.
A generalization to nonlinear magneto- and thermal
transport properties is possible.
Our scheme thus constitutes a convenient analytic way of characterizing nano-structured devices in terms
of renormalized parameters $\tilde{u}$, $\tilde{\epsilon}_d$ and $\kappa$ and the low-energy scale 
$\tilde{\Delta}$ of an underlying model. 
With the current interest in strongly correlated systems away from equilibrium our approach should prove useful 
as it provides controlled results against which more general schemes~\cite{Pletyukhov.12} might be tested.

We thank D.~Natelson, D. Schuricht, G.~Scott,  and in particular T.~Costi
for many useful discussions.
E.M. and S.K. acknowledge support by the Comisi\'{o}n Nacional de Investigaci\'{o}n Cient\'{i}fica y Tecnol\'{o}gica (CONICYT), grant No. 11100064 and the German Academic Exchange Service (DAAD) under grant No. 52636698.

{\em Note added} After completion of this work  we became aware of Ref.~\cite{Aligia.11}, which addresses the effect of p-h asymmetry on $\alpha$ within a perturbation theory around the p-h asymmetric case.
A problem with this approach is that it fails to recover p-h symmetry at $\tilde{u}=1$
and gives a linear in T term in the spectral density
away from half filling $n=1$ in contradiction to certain Ward identities~\cite{Oguri.01}.



\begin{thebibliography}{36}
\bibitem[{not({\natexlab{a}})}]{note2}
\bibinfo{note}{P-h asymmetry in $\rho_\lambda(\omega)$ will have a similar
  effect as the local p-h asymmetry generated by a finite $E_{d}$.}
\bibitem[{not({\natexlab{b}})}]{note1}
\bibinfo{note}{Note that the use of Ward identities in~\cite{Oguri.01} would
  result in a current conserving approximation up to order $V^3$ even away from
  p-h symmetry but the derivative of the (renormalized) susceptibilities with
  respect to (the renormalized) $E_d$ is not known~\cite{Yamada.79}.}
\bibitem[{not({\natexlab{d}})}]{note4}
\bibinfo{note}{We have chosen the chemical potentials to satisfy
  $\Gamma_{L}|\mu_{L}| = \Gamma_{R}|\mu_{R}|$ for convenience. Results for the
  general case are available~\cite{note3}}.
\end{thebibliography}
\end{document}